\documentclass{sig-alternate}
\usepackage[utf8]{inputenc}
\usepackage{graphicx}
\usepackage{listings}
\usepackage{fancyvrb}
\usepackage{amssymb}
\usepackage{amsmath}
\usepackage[colorlinks,urlcolor=blue,linkcolor=magenta,citecolor=red,linktocpage=true]{hyperref}

\newcommand{\MANCOOSI}{\textsc{Mancoosi}}
\newcommand{\FOSS}{\textsc{FOSS}}
\newcommand{\MOO}{\textsc{MooML}}
\newcommand{\CUDF}{\textrm{CUDF}}

\newtheorem{example}{Example}

\newcommand{\PROP}[1]{\mbox{\texttt{#1}}}

\newcommand{\SYMB}[1]{\ensuremath{\mathtt{'#1}}}
\newcommand{\KEYWORD}[1]{\ensuremath{\mathtt{#1}}}
\newcommand{\LABEL}[1]{\emph{#1}}

\newcommand{\IN}{\KEYWORD{in}}
\newcommand{\LET}{\KEYWORD{let}}
\newcommand{\MATCH}{\KEYWORD{match}}

\newcommand{\WITH}{\KEYWORD{with}}

\newcommand{\WILDCARD}{*}

\newtheorem{definition}{Definition}

\lstdefinestyle{mooml}		
               {basicstyle=\small\normalfont\ttfamily,
                 language=[Objective]Caml,
                 showstringspaces=false,
                 emph={as,let,in,fun,match,with,Unit,List,property,maximize,minimize,constraint},
                 keywordstyle=\color{black}\bf,
                 emphstyle=\bf,
               }
\lstdefinestyle{debctrl}	
               {basicstyle=\small\normalfont\ttfamily,
                 showstringspaces=false,
                 emph={Package,Version,Build,Depends,Conflicts,Provides},
                 keywordstyle=\color{black}\bf,
                 emphstyle=\bf,
               }
\lstdefinestyle{cudf}	
               {basicstyle=\small\normalfont\ttfamily,
                 showstringspaces=false,
                 emph={package,version,depends,conflicts,provides,preamble,installed,request,install,upgrade,property},
                 keywordstyle=\color{black}\bf,
                 emphstyle=\bf,
               }
\newcommand{\logsize}{\footnotesize}
\newcommand{\logwidth}{0.95\columnwidth}

\begin{document}
\conferenceinfo{IWOCE'09,} {August 24, 2009, Amsterdam, The Netherlands.} 
\CopyrightYear{2009}
\crdata{978-1-60558-677-9/09/08} 

\title{Expressing Advanced User Preferences \\ in Component
  Installation\titlenote{Partially supported by they European
    Community's 7th Framework Programme (FP7/2007--2013),
    \href{http://www.mancoosi.org}{\MANCOOSI} project, grant agreement
    n. 214898.}}

\numberofauthors{2}
\author{
  \alignauthor Ralf Treinen\\
    \affaddr{Universit\'e Paris Diderot}\\
    \affaddr{PPS, UMR 7126, France}\\
    \email{Ralf.Treinen@pps.jussieu.fr}
  \alignauthor Stefano Zacchiroli\\
    \affaddr{Universit\'e Paris Diderot}\\
    \affaddr{PPS, UMR 7126, France}\\
    \email{zack@pps.jussieu.fr}
}

\maketitle
\begin{abstract}
  State of the art component-based \emph{software collections}---such
  as \FOSS{} distributions---are made of up to dozens of thousands
  components, with complex inter-dependencies and conflicts. Given a
  particular installation of such a system, each \emph{request} to
  alter the set of installed components has potentially (too) many
  satisfying answers.

  We present an architecture that allows to express advanced user
  preferences about package selection in \FOSS{} distributions. The
  architecture is composed by a distribution-independent format for
  describing available and installed packages called CUDF (Common
  Upgradeability Description Format), and a foundational language
  called \MOO{} to specify optimization criteria. We present the
  syntax and semantics of CUDF and \MOO, and discuss the partial
  evaluation mechanism of \MOO{} which allows to gain efficiency in
  package dependency solvers.
\end{abstract}


\section{Introduction}
\label{sec:intro}

One of the noteworthy characteristics of \FOSS{} (for Free and Open
Source Software) distributions---such as Debian GNU/Linux, Red Hat
Enterprise Linux, or FreeBSD---is the availability of large numbers of
components (usually called \emph{packages} in this environment) that
can be installed, removed, and upgraded as single entities. Systems
like Debian can have up to dozens of thousands components, growing
steadily across releases and linked by complex
inter-dependencies~\cite{strongdeps-esem-2009}. Similar architectures
exist in other contexts where components are used to define
the granularity at which software can be deployed: the analogous of
\FOSS{} packages can be found for example in the Eclipse
\cite{eclipse09iwoce} and
Maven\footnote{\url{http://maven.apache.org/}} platforms; in both
cases the number of components and their inter-relationships are
similar to what exists in common \FOSS{} distributions.

In all such scenarios, user installations are managed using tools such
as \emph{package managers} which receive user requests to change the
installation in some way---e.g. install a new component---and try to
satisfy them equipped with the knowledge of where to find components
and which are their inter-relationships. When the number of components
grows, a given user request can have thousands of acceptable
solutions. For instance, in satisfying the simple ``install
wordpress'' request a package manager can be faced with questions like:
``which version of wordpress should be installed?'', ``using which web
server?'', ``relying on which PHP implementation'', etc. The number of
potential solutions for the final user can easily grow exponentially;
currently, the actual choice depends on internal heuristics
implemented by specific package managers and is customizable in ad-hoc
ways.

This paper focuses on \FOSS{} distributions and presents an
architecture to specify advanced user preferences in that context,
abstracting over package manager specific details. The architecture is
composed by two parts: a format to describe upgrade scenarios called
\CUDF{} (Common Upgradeability Description Format) and a foundational
language to encode user preferences called \MOO{} (MancOosi
Optimization Meta-Language).

The \MOO{} language is \emph{foundational} in the sense that it is not
(necessarily) meant to be a language for the end user or the system
administrator; it is rather meant as an intermediate language with a
precise semantics, which can be used by developers of installation
tools as an abstract input language for expressing user preferences,
and which on the other hand can be the target language for
representing the choice a user may have expressed, for instance using
some graphic interface. \MANCOOSI, which gives the name to \MOO, is an
ongoing project which aims, among others, to develop better algorithms
and tools to plan upgrade paths based on various information sources
about software packages and on optimization
criteria~\cite{mancoosi-wp1d1}.

\paragraph{Paper structure} The remainder of this section outlines the
upgrade process \FOSS{} packages are subject
to. Section~\ref{sec:scenario} presents common optimization criteria
for package upgrade scenarios; there criteria will serve as running
examples throughout the paper. Section~\ref{sec:cudf} summarizes
essential features of the CUDF language for describing upgrade
scenarios. \MOO{} itself and its partial evaluation mechanism are
presented respectively in Section~\ref{sec:mooml}
and~\ref{sec:partial}.

\subsection{FOSS Package Upgrade Generalities}

\begin{figure}[ht]
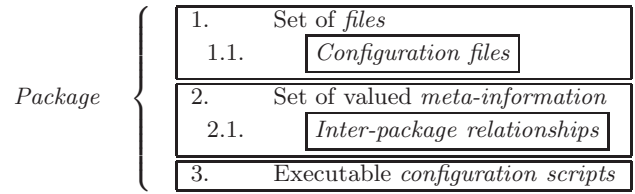

\[\mathit{Package} \quad \left\{ \quad
\begin{tabular}{|ll|}
  \hline
  1. & Set of \emph{files} \\
  ~ 1.1. & ~ ~ \fbox{\emph{Configuration files}} \\[1ex]
  \hline\hline
  2. &  Set of valued \emph{meta-information} \\
  ~ 2.1. & ~ ~ \fbox{\emph{Inter-package relationships}} \\[1ex]
  \hline\hline
  3. &  Executable \emph{configuration scripts} \\
  \hline
\end{tabular} \right.\]
\caption{\label{fig:package}Constituents of a package.}
\end{figure}

\paragraph{Packages}
\FOSS{} (binary) distributions are organized as collections of
\emph{packages}, i.e. abstractions defining the granularity at which
users can act (add, remove, upgrade, etc.) on installed software.
Abstracting over format-specific details, a \emph{package} is a bundle
of the 3 parts depicted in Figure~\ref{fig:package}.

The set of files (1) represents what the package is delivering:
executable binaries, data, documentation, etc. This set includes
configuration files (1.1), that affect the runtime behavior of the
package and are meant to be locally customized. Package
meta-information (2) contains information varying from distribution to
distribution. A common core provides: its name (a unique identifier),
a version (taken from a totally ordered set), maintainer and package
description, and most notably \emph{inter-package relationships}
(2.1). The kinds of relationship vary with the package manager used,
but there exists a de facto common subset including dependencies (the
need of other packages to work properly), conflicts (the inability of
being co-installed with other packages), feature provisions (the
ability to declare named features as provided by a given package, so
that other packages can depend on them), and restricted boolean
combinations of them~\cite{edos-package-management}. Finally, packages
come with a set of executable configuration (or \emph{maintainer})
scripts (3). Their purpose is to let package maintainers attach
actions to hooks executed by the installer; actions are used to
finalize package configuration during deployment.

\paragraph{Upgrades}
A \emph{distribution} is a collection of packages. The subset of a
distribution corresponding to the packages actually installed on a
machine is called \emph{package status} and is meant to be altered
using a \emph{package manager}. An \emph{upgrade scenario} is the
situation in which a user, typically the system administrator, submits
a \emph{user request} to the package manager, with the intention to
alter the packages status. Several entities and problems are involved
in, and should be grasped by a complete description of, an upgrade
scenario~\cite{hotswup:package-upgrades}. The main entities are
packages and the most relevant problem for the present paper is
\emph{upgrade planning}; both are briefly described below.

\begin{table}[t]
 \begin{tabular}{l}
  \begin{minipage}[c]{\logwidth}
   \begin{Verbatim}[commandchars=\\\{\},fontsize=\logsize]
# apt-get install aterm 
   \end{Verbatim}
  \end{minipage}
  \\\hline
  \begin{minipage}[c]{\logwidth}
   \begin{Verbatim}[commandchars=\\\{\},fontsize=\logsize]
Reading package lists... Done
Building dependency tree... Done
The following extra packages will be installed:
  libafterimage0
0 upgraded, 2 newly installed, 0 to remove and
  1786 not upgraded.
Need to get 386kB of archives.
807kB of additional disk space will be used.
   \end{Verbatim}
  \end{minipage}
  \\\hline
  \begin{minipage}[c]{\logwidth}
   \begin{Verbatim}[commandchars=\\\{\},fontsize=\logsize]
Get: 1 http://ftp.debian.org libafterimage0 2.2.8
Get: 2 http://ftp.debian.org aterm 1.0.1-4
Fetched 386kB in 0s (410kB/s)
   \end{Verbatim}
  \end{minipage}
  \\\hline
  \begin{minipage}[c]{\logwidth}
   \begin{Verbatim}[commandchars=\\\{\},fontsize=\logsize]
Selecting package libafterimage0.
(Reading database ... 294774 files and dirs ...)
Unpacking libafterimage0 ...
Selecting package aterm.
Unpacking aterm (aterm_1.0.1-4_i386.deb) ...
   \end{Verbatim}
  \end{minipage}
  \\\hline
  \begin{minipage}[c]{\logwidth}
   \begin{Verbatim}[commandchars=\\\{\},fontsize=\logsize]
Setting up libafterimage0 (2.2.8-2) ...
Setting up aterm (1.0.1-4) ...
   \end{Verbatim}
  \end{minipage}
 \end{tabular}

 \caption{The package upgrade process. Horizontal lines separate the
   phases described in the text.}
 \label{tab:upgrade-phases}
\end{table}

Table~\ref{tab:upgrade-phases} summarizes the different phases of the
\emph{upgrade process}, using as an example the popular
\texttt{apt-get} package manager (others follow a similar process).
Phase (1) is a user specification of how the package status should be
altered. The expressiveness of the request language varies with the
package manager: it can be as simple as requesting the
installation/removal of a single package by name, or can also enable
limited expression of per-package preferences such as APT
pinning~\cite{apt-howto}. Phase (2) (dependency resolution) checks
whether a package status satisfying all dependencies and user request
exists, it has been shown that this problem is at least
NP-complete~\cite{edos-package-management}. If this is the case, one
such package status is chosen---trying to satisfy user preferences, if
any---and gets called \emph{solution}. Deploying a new status
corresponding to the solution consists of package retrieval (3) and
unpacking (4), possibly intertwined with several configuration phases
(5) where maintainer scripts get executed.

Various challenges related to the upgrade process still need to be
properly addressed. An example of a very practical challenge is the
need to provide \emph{transactional upgrades}, offering the
possibility to roll back in case an unexpected (and unpredictable in
general) failure is encountered during upgrade
deployment~\cite{hotswup:package-upgrades}. Other challenges concern
\emph{upgrade planning}. For instance, dependency resolution can fail
either because the user request is unsatisfiable (e.g., user error or
inconsistent distributions~\cite{edos2006ase}) or because the package
manager is unable to find a solution. Completeness---the guarantee
that a solution will be found whenever one exists---is a desirable
package manager property~\cite{mancoosi-debconf8}, unfortunately
missing in most package managers, with too few claimed
exceptions~\cite{niemeyer-smart,tucker-opium}.

\paragraph{User Preferences}
While suitable and complete techniques to provide dependency solving
completeness are now well-known~\cite{edos2006ase} and ``just'' lack
widespread adoption, \emph{handling of complex user preferences} is a
novel problem for software upgrade, and is the main concern of this
paper. It boils down to let users specify what constitutes the
``best'' solution among all acceptable solutions, and provide
mechanisms to efficiently find it. Example of preferences are
\emph{policies}~\cite{niemeyer-smart,trezentos07}, like minimizing the
download size or prioritizing popular packages, and also more specific
requirements such as blacklisting packages maintained by an untrusted
maintainer.

The first necessary step to attack the problem is devising a way to
encode user preferences in a flexible way, without hindering package
manager ability to respect them. A prerequisite of that is a rigorous
description of upgrade scenarios, on top of which the \emph{meaning}
of user preferences will be defined.


\section{User preference scenarios}
\label{sec:scenario}

In the following we will consider several possible scenarios where
user needs can be better encoded as user preferences in \MOO. The
actual encoding in \MOO{} will be presented in
Section~\ref{sec:examples:mooml}, after a more in depth presentation
of the language.

\begin{description}

\item[Size] Minimizing the total size consumed by the package
  installation is a rather most basic optimization criterion and a
  frequent need of package managers for embedded systems.

\item[Freshness] Preferring more recent package versions
  over \linebreak older package versions is also very common, and
  hard-wired in most package managers. The hard-wiring in Debian's APT
  as a \emph{hard} constraint is the main cause for the incompleteness
  of its dependency solving abilities.

\item[Pinning] To avoid forcing the choice of the most recent version
  of a package in all cases, APT enables to specify different choices
  for specific packages by the mean of a mechanism called
  \emph{pinning}~\cite{apt-howto}. In its essence, pinning consists in
  specifying integer score values (called \emph{priorities}) for
  individual packages based on patterns of package names, package
  versions, and origin; among all the versions of a given package, the
  one with the highest priority gets chosen. By default, priority
  follows versions. This is an example of ``local'' preferences that
  apply to particular packages, in contrast to uniform constraint
  like total installation size.

\item[Security updates] usually should have highest priority\linebreak
  while choosing which packages have to be upgraded. We will
  demonstrate \MOO's multi-criteria capabilities by stating that
  maximizing security updates has priority over package freshness.

\item[Multiple packages] Some package managers, most notably
  \texttt{rpm}, allow for multiple versions of the same package to be
  installed; while this is an interesting property, one might want to
  automatically ``clean up'' useless multiple installations. This
  scenario will show how to minimize the number of packages that are
  installed in multiple versions.

\end{description}

Note that, while they are presented as such for the sake of brevity,
scenarios are not mutually exclusive in practice. In our vision, some
optimization criteria will constitute a default configuration of a
given package manager (e.g.: always prioritizing security upgrades,
avoiding package downgrades, etc.) while some other will be added by
users by the mean of specific user interfaces. Even when the latter
possibility is not exploited, there are advantages in externalizing
preferences which are currently hard-wired in solving algorithms: for
instance they will become overwritable by users and it will be easier
to share optimizers among distributions.

\MOO{} allows to combine multiple optimization criteria, however one
has to specify a hierarchy among the multiple critera. For instance
one can require to search for a solution that is minimal in size
first, and among all that solutions that are minimal in size to choose
one with maximal freshness. It is not possible to optimize two
independent criteria at the same time since in that case an optimal
solution might not exist.

As next section will explain, optimization criteria do not allow to
taint the correctness of a solution, e.g. by allowing to install at
the same time two conflicting packages.

\section{Describing Upgrade Scenarios}
\label{sec:cudf}

State of the art mechanisms for specifying user preferences
highlighted so far~\cite{niemeyer-smart,tucker-opium,apt-howto} suffer
from two main drawbacks: they are package manager specific, and they
are not expressive enough to encode all our user preference scenarios
(see Section~\ref{sec:scenario}). The first step we pursue in
addressing these shortcoming is devising a rigorous format in which
upgrade scenarios can be encoded; a user preference language will be
then developed on top of such a format (see Section
Section~\ref{sec:mooml}). The format is called CUDF (Common
Upgradeability Description Format).  The specification of CUDF
\cite{mancoosi-wp5d1} had been guided by some general \emph{design
  principles}:

\begin{description}

\item[Be distribution agnostic] One of the main purposes of CUDF is
  being a \emph{common} format to encode upgrade scenarios coming from
  heterogeneous environments. \linebreak As a consequence, CUDF is
  agnostic to distribution specific details such as the used package
  system or package manager.

\item[Stay close to the original problem] While there are
  several possible encoding of upgrade scenarios~\cite{edos2006ase},
   CUDF aims to be as close as possible to the original
  problem, in order to preserve the ability for humans to understand
  the pre-CUDF upgrade scenario, and ease interoperability with legacy
  package managers.

\item[Extensibility] Core package properties---e.g.: name, version,
  dependencies, \ldots---are shared by all distributions and essential
  to grasp the meaning of upgrade scenarios. Other auxiliary
  properties are not, but might be the subject of user preferences
  (e.g., minimize the number of ``buggy'' packages, according to
  distribution specific buggyness notions). In order not to hinder the
  possibility to express such user preferences on top of CUDF, the
  format allows to specify \emph{extra package properties} not
  prescribed by the format specifications.

\item[Transactional semantics] The point of view of CUDF is upgrade
  planning: the notion of correctness of a solution with respect to an
  upgrade scenario expressed in CUDF is global and does not express
  the package deployment steps needed to pass from the starting
  package status to the final one. Such steps are more low-level, and
  mostly uninteresting for user preferences.

\item[Plain text format] Technically, CUDF aims at being simple to
  parse and to generate. The reason is the consciousness of the
  generality of the user preference problem and the desire to make the
  format popular among different distributions. As plain text is the
  universal encoding for information interchange formats in \FOSS{}
  communities~\cite{esr03artofunix}, using a plain text format makes
  it easy for package manager developers to adapt tools to CUDF.

\end{description}

\subsection{CUDF Syntax}

\begin{figure}[ht]
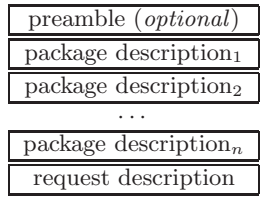

  \begin{center}
    \begin{tabular}{|c|}
      \hline
      preamble (\emph{optional}) \\
      \hline\hline
      package description${}_1$ \\
      \hline\hline
      package description${}_2$ \\
      \hline
      \multicolumn{1}{c}{$\cdots$} \\
      \hline
      package description${}_n$ \\
      \hline\hline
      request description \\
      \hline
    \end{tabular}
  \end{center}
  \caption{Overall structure of a CUDF document.}
  \label{fig:cudf-structure}
\end{figure}

The CUDF encoding of an upgrade scenario assumes the name of
\emph{CUDF document}. Every such document has an abstract logical
structure, a formal meaning, and a serialized form as a plain text
file. The \emph{logical structure} of a CUDF document---sketched in
Figure~\ref{fig:cudf-structure}---is based on \emph{stanzas}, which
are collections of key-value pairs called \emph{properties}. Values
are typed within a simple \emph{type system} containing basic data
types (e.g.: integers, boolean, and strings) and more complex,
package-specific, data types such as boolean formulae over versioned
packages used to represent inter-package dependencies.

CUDF documents contain one \emph{package description stanza} for each
package known to the package manager; collectively they represent the
\emph{package universe}. This means that both installed and
non-installed (but available) packages are represented in the same way
in the same document, in contrast to current package installation
systems which often distribute this information over different files
using different syntactic representations.

Package description stanzas are based on a core set of properties
(sometime optional, but always with default values), the most
important of which are: \PROP{package} and \PROP{version} (which
unambiguously identify packages), \PROP{depends} and \texttt{conflicts}
(which express package dependencies and conflicts to be properly
installed), \PROP{provides} (which expresses versioned \emph{features}
that the current package provides for other packages to depend or
conflict upon), and \PROP{installed} (which state whether the current
package is installed or not).

\begin{figure}[tbh]
  \begin{lstlisting}[style=cudf]
preamble: 
property: suite: enum(stable,unstable) = \
 "stable"
property: bugs: int = 0

package: car
version: 1
depends: engine, wheel, door, battery
installed: true
bugs: 183

package: bicycle
version: 7
suite: unstable

package: gasoline-engine
version: 1
depends: turbo
provides: engine
conflicts: engine, gasoline-engine
installed: true

...

request:
install: bicycle, gasoline-engine = 1
upgrade: door, wheel > 2
  \end{lstlisting}
  \caption{Sample CUDF document.}
  \label{fig:cudf-example}
\end{figure}

Figure~\ref{fig:cudf-example} shows the serialization of a sample CUDF
document. As stanzas are separated by blank lines, the central part of
the figure shows three package description stanzas, starting with the
\PROP{package} property, where both core and extra properties are
used. The latter must be declared in the optional \emph{preamble
  stanza}, which starts the document in
Figure~\ref{fig:cudf-example}. The ability to declare extra properties
accounts for extensibility and also enables to statically verify the
syntactic correctness of CUDF documents. The bottom part of
Figure~\ref{fig:cudf-example} shows the \emph{request description
  stanza}, where the user request is expressed. In its minimal form,
such stanza is used to express which packages the user wants to
\PROP{install}, \PROP{remove}, or \PROP{upgrade} (using the homonym
properties), possibly specifying version requirements.

The example lacks the encoding of user preferences. This lack, which
was our initial motivation for the work reported here, can be filled
by an optional property specifiable in the request stanza, called
\PROP{preferences}. Its content is a \MOO{} program, discussed in the
next section. What is relevant here is that \MOO{} programs may be
part of CUDF documents and will be able to express preferences
referencing CUDF stanzas.

\subsection{CUDF Semantics}

Given that a CUDF document completely describes an upgrade scenario,
what does constitute its \emph{meaning}? Intuitively, an upgrade
scenario poses a challenge for the package manager, its solutions are
new package statuses. The meaning, or semantics, of a CUDF document is
hence a characterization of all \emph{valid solutions} matching the
upgrade scenario. We recall that a package status is just a set of
packages contained in the package universe which we know is fully
encoded in the document. On that basis, we declare that a solution is
valid if and only if:
\begin{enumerate}
\item all installed packages have their dependencies satisfied,
  i.e. installed as well (\emph{abundance});
\item no two packages that are in conflict are installed together
  (\emph{peace});
\item the user request is satisfied by installed packages
  (\emph{correctness}).
\end{enumerate}

The first two points have been previously formalized relying on an
encoding in propositional logics~\cite{edos2006ase}. That encoding
fails to respect the design principle of staying close to the original
problem since, for example, packages with the same name and different
versions are treated as unrelated boolean variables in the
encoding. The formal semantics of CUDF characterizes all valid
solution corresponding to a given CUDF document as a binary relation
among package statuses, indexed by the user request. We will not give
the full details here, for which the reader is referred
to~\cite{mancoosi-wp5d1}, but rather only discuss the
\emph{peculiarities} of CUDF formal semantics and its differences with
respect to previous encodings.


An important semantic difference between existing package management
systems in \FOSS{} distributions is whether they a priori allow
packages to be installed in multiple versions (like \texttt{rpm} does)
or not (like \texttt{dpkg}). CUDF semantics here follows the
\texttt{rpm} philosophy of a priori allowing multiple versions of a
package to be installed at the same time. To encode Debian-like
upgrade scenarios, where different versions of the same package are
forcibly in conflict, a special case of conflicts semantics is
exploited, namely: self-conflicts are ignored. Hence, in
Figure~\ref{fig:cudf-example}, all packages (potentially) appearing in
multiple versions declare an (unversioned) conflict with themselves,
as it happens for \texttt{gasoline-engine}. The semantics ensures that
such conflicts are ignored for the very same version of the package
(otherwise those packages will be useless) but take effect on
different versions of \texttt{gasoline-engine}, granting that only one
version of it can be installed. Such a semantics is coherent with
self-conflicts on virtual packages, which can be exploited to ensure
mutual exclusions among different providers of the same feature. For
instance, three packages like \texttt{postfix}, \texttt{sendmail}, and
\texttt{qmail}, all providing the \texttt{mail-transport-agent}
feature, can be made mutually exclusive by having all of them both
provide \emph{and} conflict with \texttt{mail-transport-agent}.

Finally, feature provision via \PROP{provides} is versioned, meaning
that specific versions of a given feature can be provided. Not
specifying a version---as in \texttt{provides:~foo}---is interpreted
as providing \emph{all versions} of the \texttt{foo} feature.

Equipped with all this, verifying the satisfaction of a user request
boils down to re-use the notions of peace and abundance: an
\PROP{install} request is satisfied if and only if the same line,
considered as a dependency, would be satisfied (abundance); a
\PROP{remove} request is satisfied if and only if a corresponding
conflict is unsatisfied (peace). Only \PROP{upgrade} needs some
caution: in principle it can be handled as \PROP{install}, but
additionally it also requires that all packages mentioned in the user
request are installed in a single version. Furthermore, after
\PROP{upgrade}-ing some package we must have a version of that package
that is at least as new as any previously installed version of that
package.

CUDF also allows to express that a particular package must not be
removed, that it must be kept in its current version, or that its
functionality must be provided by some package (see
\cite{mancoosi-wp5d1} for details).

\subsection{CUDF Implementations}

CUDF has already seen various implementations. The first
implementation---\texttt{libcudf}---is the ``reference''
implementation of the CUDF specifications and has been developed by
one of this paper authors. \texttt{libcudf} consists in a library able
not only to parse and pretty print CUDF documents, but also to verify
the CUDF semantics. This latter feature can be exploited in two ways:
\begin{enumerate}
\item given a CUDF document, \texttt{libcudf} can verify whether the
  contained package status is \emph{consistent}, i.e., whether
  abundance and peace are verified for all its packages;

\item given a CUDF document and an encoding of a potential solution,
  \texttt{libcudf} can verify whether the solution is valid, i.e.,
  abundance + peace + user-request satisfaction.
\end{enumerate}

\texttt{libcudf} comes with the \texttt{cudf-check} command line tool
 which provides the above two features out of the box.  The
library is Free Software and can be user both from the OCaml and C
programming languages; it is available for download at
\url{http://www.mancoosi.org/software/}.

The authors are aware of other CUDF implementations. Some of them are
being developed within \MANCOOSI{} to convert distribution-specific
upgrade scenario descriptions into CUDF, so that a cross-distribution
corpus of upgrade scenarios can be formed. They will be released
shortly at least for the following distributions: Mandriva,
CaixaMagica, Debian GNU/Linux. Using such tools we have verified that
the average size of an upgrade scenario encoded in CUDF is linear with
the size of the origin package manager information and usually
smaller.\footnote{e.g. on a large Debian installation, using both
  testing and unstable package repositories for about 45'000 packages,
  the package manager information on disk amounts to 14 Mb and the
  corresponding CUDF document has 9 Mb.}

Another independent implementation is already available in
CUPT\footnote{\url{http://wiki.debian.org/Cupt}}, a new APT-compatible
package manager for Debian. In CUPT, CUDF is used as a syntactic
format to pipe upgrade scenarios to external solvers, so that upgrade
planning can be decoupled from other package manager activities. Also,
such a choice enables sharing more easily dependency solvers not only
inter- and intra-distributions, but also with the scientific
community.


\section{Expressing User Preferences}
\label{sec:mooml}

Having a rigorous description of upgrade scenarios, we can now devise
our language to express user preferences. Our proposal for such a
language---\MOO{} for MancOosi Optimization Meta\footnote{The
  \emph{meta} is inherited from the ML family of languages, for our
  purpose there is no distinguished meta level.}-Language---is
described in this section. The design of the language needs to face
two requirements that appear to be in mutual conflict:
\begin{description}
\item[Simplicity] programs written in \MOO{} have to be interpreted by
  solver tools that will try to satisfy user preferences. Hence they
  should be as simple as possible in order to minimize the burden put
  on the developers of these tools.

\item[Expressivity] the \MOO{} language should allow to express
  sophisticated optimization criteria expressive\linebreak enough to
  encode the scenario we have discussed.

\end{description}
The right choice of a language was to be found between two
extremes. On one extreme a \emph{Turing-complete programming language}
with rich user-defined data structures and function definitions
through unrestricted recursion. This extreme would provide maximum
expressivity by definition, but would require tool developers to
integrate an interpreter for a full-fledged programming language. On
the other extreme a \emph{restricted language} allowing only for
simple combinations of optimization criteria for which a limited
choice of common simple criteria is provided. This extreme would
probably make life easy for tool implementers, but would be too
limited in expressivity. It would also bear the risk of being obliged
to continuously extend the choice of optimization criteria.

In order to find the right balance between these two extremes we made
the following design choices for \MOO:
\begin{itemize}
\item \MOO{} allows to separately specify hard constraints that must
  be satisfied by ``user-approved'' solution, and optimization
  criteria.

\item \MOO{} does not allow to program directly an algorithm that
  compares alternative solutions\footnote{as it happens, for example,
    with the \texttt{sort} function provided by the standard library
    of several programming languages}. Instead, the language allows to
  define how to compute a \textit{measure} of solution quality. Two
  possible solutions are compared by comparing their respective
  measures.  A \MOO{} program specifies the polarity of each measure
  (i.e., whether it should be minimised or maximised). In case several
  measures are defined the program defines a strict priority hierarchy
  (technically this is a \emph{lexicographic combination} of orders).

\item \MOO{} is a strongly typed functional language allowing for
  polymorphic types and inference of principal types.

\item \MOO{} does not allow for arbitrary use of recursion, and is
  deliberately not Turing complete. Instead it provides for a generic
  \texttt{fold}-like iterator over lists, which allows to program
  primitive recursive functions over lists.

\item \MOO{} does not allow to define custom data types.

\item \MOO{} does not have a mechanism to catch exceptions but allows
  to express execution errors.

\end{itemize}

\subsection{MooML Programs}

\begin{figure}
  \[\begin{array}{rcl@{\hspace{1em}}r}
  P & ::= & & \LABEL{\textbf{program}} \\
  & & (~\mathtt{\LET~x=e}~)* & \LABEL{definition} \\
  & & (~\mbox{\texttt{constraint~e}}~)? & \LABEL{constraint} \\
  & & (~(\mbox{\texttt{minimize}} ~|~ \mbox{\texttt{maximize}})
    ~\mathtt{e}~)*
    & \LABEL{criteria} \\
  \end{array}\]
  \caption{Syntax of \MOO{} programs}
  \label{fig:moo-structure}
\end{figure}

The high-level syntax and structure of a \MOO{} program is sketched in
Figure~\ref{fig:moo-structure}. Such a program is composed by a series
of preparatory global definitions, meant to be reused in the remainder
of the program. Then, two main parts compose a \MOO{} program. The
first is a \emph{constraint}, that is a boolean expression which, when
evaluated to \texttt{true}, indicates a solution considered acceptable
by the user. Using a constraint users can exclude solutions that, in
spite of being valid with respect to CUDF semantics, are undesirable
for them. The second part is a list of \emph{optimization criteria},
i.e. expressions of the language returning integers and tagged with a
request to either minimize or maximize them over all otherwise valid
solutions.

\begin{figure}[t!]
  \[\begin{array}{rcl@{\hspace{1em}}r}
  e & ::= & & \LABEL{\textbf{expressions}} \\
  & & \mathtt{x} & \LABEL{variable} \\
  & | & \mathbb{C}_v & \LABEL{literal} \\
  & | & \mbox{\texttt{fun x -> e}} & \LABEL{abstraction} \\
  & | & \mathtt{e ~ e} & \LABEL{application} \\
  & | & \mathtt{()} & \LABEL{unit} \\
  & | & \mathtt{(e_1,\ldots,e_n)} & \LABEL{tuple} \\
  & | & \mathtt{\{l_1=e_1,\ldots,l_n=e_n\}} & \LABEL{record} \\
  & | & \mathtt{[~]} & \LABEL{empty list} \\
  & | & \mathtt{e~::~e} & \LABEL{list} \\
  & | & \mathtt{e.l} & \LABEL{projection} \\
  & | & \mathtt{\LET~p=e_1~\IN~e_2} & \LABEL{let binding} \\
  & | & \begin{array}[t]{l}
    \mathtt{\MATCH~e~\WITH} \\
    \mathtt{~ p_i\Rightarrow e_i ~ |\cdots| ~ p_n\Rightarrow e_n} \\
  \end{array}
  & \LABEL{pat. match} \\[3ex]

  p & ::= & & \LABEL{\textbf{patterns}} \\
  & & \mathtt{x} & \LABEL{variable} \\
  & | & \mathbb{C}_v & \LABEL{constant} \\
  & | & \mathtt{()} & \LABEL{unit} \\
  & | & \mathtt{(p_1,}\ldots, \mathtt{p_n)} & \LABEL{tuple} \\
  & | & \mathtt{\{l_1=p_1,\ldots,l_n=p_n\}}
  & \LABEL{record} \\
  & | & \SYMB{x} & \LABEL{enumeration} \\
  & | & \mathtt{[~]} & \LABEL{empty list} \\
  & | & \mathtt{p_1 :: p_2} & \LABEL{list} \\
  & | & \mathtt{\WILDCARD} & \LABEL{wildcard} \\[3ex]

  \mathbb{C}_v & ::= & & \LABEL{\textbf{CUDF literals}} \\
  & & \mbox{\texttt{true}} ~|~ \mbox{\texttt{false}} & \LABEL{booleans} \\
  & | & \ldots ~|~ \mbox{\texttt{-1}} ~|~ \mbox{\texttt{0}}
        ~|~ \mbox{\texttt{1}} ~|~ \ldots & \LABEL{integers} \\
  & | & \mbox{\texttt{"s"}} & \LABEL{strings} \\
  & | & \mbox{\texttt{'l}} & \LABEL{enumerations} \\
  & | & \ldots & \LABEL{formulae}, \ldots \\
  \end{array}\]

  \caption{Syntax of \MOO{} expressions}
  \label{fig:moo-syntax}
\end{figure}

The syntax of \MOO{} \emph{expressions}, as given in
Figure~\ref{fig:moo-syntax}, has features borrowed from common
functional programming languages. Expressions sport rich types such as
records, tuples and lists defined on top of the basic CUDF types, as
well as expressive constructs such as pattern matching and
(non-recursive) local definitions. The evaluation of a \MOO{} program
is a straightforward ML-style evaluation~\cite{kahn-miniml} with
pattern matching~\cite{augustsson-patternmatching}; overall it boils
down to evaluate the constraint and optimization criteria expressions
in an evaluation environment enriched with global
definitions. Additionally, the environment is also enriched with:
\begin{itemize}
\item the \MOO{} \emph{standard library}, which provides the usual kit
  of functional programming functions and in particular the
  \texttt{fold} iterator (and some of its derivatives, like
  \texttt{map}, and \texttt{filter}) without which iterating over
  list data structures would be impossible within the language;
\item the \emph{package universe} \texttt{u} denotes a list of records
  representing all the package description stanzas of the CUDF
  document from which the \MOO{} program originated. Each record
  contains one field for each package property, and can therefore be
  properly typed having around the CUDF preamble. However, the
  \PROP{installed} property gets split into two new properties:
  \begin{description}
  \item[\PROP{was-installed}] (the same as the original
    \PROP{installed}, renamed for clarity) denotes whether the owning
    package \emph{was installed} in the upgrade scenario presented to
    the package manager
  \item[\PROP{is-installed}] denotes whether the owning package
    \emph{is installed} in the proposed solution in the context of
    which the \MOO{} program is being evaluated
  \end{description}
\item the \emph{user request} \texttt{r} denotes a record corresponding to
  the user request stanza of CUDF.
\end{itemize}

The only way to express iteration over lists is to use the predefined
\texttt{fold} function. An expression
\begin{lstlisting}
  (fold f [an ; ... ; a1] a0) 
\end{lstlisting}
is evaluated as 
\begin{lstlisting}
  (f an (f a(n-1) ... (f a1 a0) ... ))
\end{lstlisting}
An alternative way to describe its semantics is the following
iterative pseudo-code:
\begin{verbatim}
  r := a0;
  foreach i in 1 .. n do r := (f ai r);
  return r;
\end{verbatim}
For instance, the standard library contains a definition of the \texttt{sum}
function to sum up a list of integers:
\begin{lstlisting}
  let sum l = fold add l 0
\end{lstlisting}
and other functions like \texttt{filter}, \texttt{map}, \texttt{max},
etc.\ acting on lists can easily be defined the same way.

Note that all properties, except the \PROP{is-installed} property, of
packages are given by the CUDF document on which a \MOO{} program is
applied. The input CUDF document describes the \PROP{was-installed}
property of any package, it is the role of the \MOO{} program to
impose constraints on the possible \PROP{is-installed} properties of
packages, and to calculate a score on any possible choice of
\PROP{is-installed} properties of the packages.

Once the constraint and criteria expressions are fully reduced, there
are enough information to know whether or not the solution
should be discarded (constraint evaluated to \texttt{false}). If it is
not the case, the different criteria values together denote a tuple
that can be \emph{lexicographically} compared with tuples coming from
other candidate solutions to determine which of the two is to be
preferred. Of course the lexicographic order should take into account
the ``polarity'' of the criterion, i.e., whether it was a
\texttt{minimize} or \texttt{maximize} request.

Types are not explicitly given in the syntax of the language, because
they can be reconstructed in the style of
Damas-Milner~\cite{milner-principal-type}, obtaining principal
types. The only source of ambiguity in the type system are record
labels which, due to the CUDF ability to declare extra properties, may
be not sufficient to unambiguously determine record types. While there
seems to be no obstacles in extending the type system to account for
them in the style of Remy~\cite{remy-typeinf-records}, we have
preferred to provide optional type ascriptions in the concrete syntax
(not shown in Figure~\ref{fig:moo-syntax}) to disambiguate the rare
ambiguous cases.

\subsection{Examples}
\label{sec:examples:mooml}

\MOO{} is expressive enough to account for all usage scenarios
presented in Section~\ref{sec:scenario}, as we will show in the
following.  The need of program simplicity for solver implementers
will be addressed by the partial evaluation mechanism in the next
section.

\begin{example}[minimize total installation size]~\\ The ``Hello,
  World!'' equivalent in \MOO{} is likely to be the widespread policy
  of minimizing the total installation size, very useful for embedded
  or otherwise constrained systems. It can be expressed as:
  \begin{lstlisting}
let size pl =
  sum (map (fun p -> p.installed-size)) pl
minimize size
  (filter (fun p -> p.is-installed) u)
\end{lstlisting}
  where \texttt{sum} is a library function summing up integers. The
  program simply states that the score to be minimized is the sum of
  the \PROP{installed-size} value (an extra property with the obvious
  meaning) of all packages installed in the proposed solution.
\end{example}

\begin{example}[maximize package ``freshness'']~\\
  \label{ex:freshness}
  The scenario requiring to maximize the number of packages installed
  at their most recent version can be expressed as follows.
  \begin{lstlisting}
let is-recent p =
  forall
   (fun q -> (q.name != p.name)
          || (q.version <= p.version)
   u
maximize cardinality
  (fun p -> p.is-installed && is-recent p) u
  \end{lstlisting}
  \texttt{is-recent} is used as an auxiliary function to check whether
  a given package---given as its record---is the most recent version
  of all equally named packages; its implementation relies on
  \texttt{forall} which check the \texttt{true}-ness of a boolean
  predicate over a list of items (in this case, the package universe
  \texttt{u}). Complementary, \texttt{cardinality} \emph{counts} the
  number of times a predicate is \texttt{true} over a list; in the
  given optimization criteria, it is used to require the maximization
  of ``recent'' packages.
\end{example}

\begin{example}[flexible APT pinning]~\\
  \label{ex:apt-pinning}
  APT pinning can be encoded in at least a couple of different ways
  using \MOO, depending on the desired goal. A first possibility is to
  encode the \emph{exact} semantics of pinning, so that the only
  acceptable solutions will be those potentially returned by a pinning
  implementation. In essence, pinning works at the package choice
  level, ensuring that among all available versions of a given
  package, the one with the highest \emph{pin priority} is
  installed. While pin priority themselves can be assigned using
  \MOO{} (see Section~\ref{sec:partial}), if we assume that each
  package comes with an extra property \PROP{pin-priority}, we can
  encode pinning semantics as follows:
  \begin{lstlisting}
let max-pin p =
  max (map (fun z -> z.pin-priority)
         (filter (fun q -> q.name == p.name) u))
constraint forall (fun p -> p.is-installed
             && p.pin-priority = max-pin p)
  \end{lstlisting}
  Given that this strict semantics is a well-known cause of APT
  incompleteness \cite{edos-package-management}, a more ``flexible''
  pinning encoded can be obtained by requiring to maximize the number
  of packages at maximal pin priority:
  \begin{lstlisting}
let max-pin p = (* as above *)
maximize cardinality
  (fun p -> p.is-installed
            && p.pin-priority = max-pin p)
  \end{lstlisting}
  An even more flexible metric over APT pinning can be obtained by
  minimizing the total difference between maximal and actual pin
  priorities as follows:
  \begin{lstlisting}
let max-pin p = (* as above *)
minimize sum
  (map (fun p -> if p.is-installed
            then max-pin p - p.pin-priority
            else 0) u)
  \end{lstlisting}
\end{example}

\begin{example}[priority to security updates]~\\
  The scenario which requires to prioritize security upgrades
  over any other criteria can be encoded straightforwardly by relying
  on \MOO's lexicographic ordering over solution measures. In the
  following example it is combined with the freshness criteria of
  Example~\ref{ex:freshness}.
  \begin{lstlisting}
let is-recent p = (* as above *)
maximize cardinality
  (fun p -> p.is-installed && not p.was-installed
            && p.is-security-fix) u
maximize cardinality
  (fun p -> p.is-installed && is-recent p) u
  \end{lstlisting}
  Note that we explicitly require the package to be newly installed
  before verifying whether it is a security fix (extra property
  \PROP{is-security-fix}), this way we ensure the security fix is
  being delivered with the proposed solution. Lexicographic ordering
  ensures that solutions with a higher number of security fixes being
  delivered will be preferred, no matter the total freshness. (How to
  improve the example to ensure that no past security fixes get
  removed by downgrades is left as an exercise.)
\end{example}

\begin{example}[minimize multiple versions]~\\
  \label{ex:multiple:versions}
  In this example we wish to minimize the number of packages that exist
  in multiple versions in the final installation.

  \begin{lstlisting}
let number-versions p = length
   (filter (fun q -> q.is-installed &&
                     p.name = q.name)
           u)
minimize cardinality
   (fun p -> p.is-installed &&
             number-versions p > 1) u
\end{lstlisting}

  The function \texttt{number-versions} applied to package $p$
  calculates the number of installed packages with the same name as
  the package $p$. We minimize the number of installed packages for
  which the function \texttt{installed-version} returns a value
  strictly greater than $1$.
\end{example}


\section{Partial evaluation}
\label{sec:partial}

\begin{figure}
  \begin{center}
    \includegraphics[width=\columnwidth]{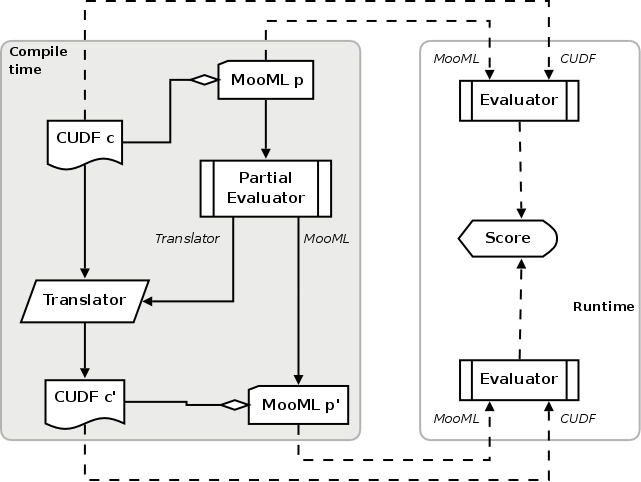}
    \caption{Partial evaluation and its properties}
    \label{fig:partial-evaluation}
  \end{center}
\end{figure}

In their full generality, \MOO{} programs can be too complex to handle
for dependency solvers, or at least require non trivial implementation
efforts to develop a full \MOO{} evaluator. To address this
shortcoming, \MOO{} has been designed to be a good subject for
\emph{partial evaluation} which processes fully general \MOO{}
programs and returns ``simpler'' programs, ideally more suitable for
digestion by dependency solvers. More precisely (see
Figure~\ref{fig:partial-evaluation}), \MOO{} partial evaluation is
applied to a program~$p$, which belongs to a CUDF document~$c$, and
returns two new entities: a \emph{new program}~$p'$ and a \emph{CUDF
  transformer} applicable to ``$c$-like'' CUDF documents, intuitively
documents sharing the same extra properties of $c$. Once applied to
$c$, the transformer returns a new document $c'$, to which $p'$
belongs. Partial evaluation enjoys the property that the evaluation of
$p$ in the context of $c$ returns the same result (constraint and
measure tuple) than the evaluation of $p'$ in the context of $c'$. The
advantage of $p'$ over $p$ is that it is potentially simpler, in the
sense that it can be implemented by ignoring significant parts of the
\MOO{} language. However, in the worst case, the partial evaluator may
not be able to do any simplification.

The guiding principle of \MOO's partial evaluation is to pre-compute
all sub-expressions that depend on the upgrade scenario, but not on
the upgrade solution, and to ``save'' them as fresh package
properties. As a consequence, $p'$ is obtained by substituting complex
sub-expressions with access to (fresh) properties, and $c'$ is
obtained by adding (fresh) properties. To characterize a little more
formally, the sub-expressions that are good partial evaluation
candidates we first define an equality relation which relates all
package statuses equal up to \PROP{is-installed}:
\begin{definition}[sibling package lists]
  Two package lists $l_1,l_2$ are \emph{siblings}, written $l_1\Bumpeq
  l_2$, if $\mathit{Dom}(l_1)=\mathit{Dom}(l_2)$ (i.e., they contain
  the same packages), and for each $(p,v)\in \mathit{Dom}(l_1)$ we
  have that $l_1(p,v)$ equals $l_2(p,v)$ except possibly the value of
  the \PROP{is-installed} property.
\end{definition}
Then, we grasp partial-evaluable (sub-)expressions with the notion of
local expressions. In the following definition we will make use of a
mathematical semantics of the \MOO{} language (the formal definition
of which is omitted from this paper). When $e$ is a \MOO{} expression
and $\sigma$ an evaluation environment mapping identifiers to semantic
values, then $[[e]]\sigma$ denotes the semantic object obtained by
evaluating $e$ in the environment $\sigma$.
\begin{definition}[local expressions]
  \label{def:local}
  An expression $e$ of type $\mathtt{package}\to \mathtt{t}$, and
  which does not have any unbound identifiers besides \(\mathtt{r}\)
  and \(\mathtt{u}\), is called \emph{local} if for all packages $p$,
  for all package lists $l_1,l_2$, request $r_0$ such that $l_1\Bumpeq
  l_2$ and $p\in l_1$, $p\in l_2$ we have that
  \[ [[e]][\mathtt{u}\mapsto l_1, \mathtt{r}\mapsto r_0](p)
  = [[e]][\mathtt{u}\mapsto l_2, \mathtt{r}\mapsto r_0](p) \]
\end{definition}
Intuitively, local expressions are all those expressions whose
evaluation does not depend on the \PROP{is-installed} values of
packages coming from the package universe; note that expressions
accessing the \PROP{is-installed} property of their argument can be
local nevertheless. As stated in Definition~\ref{def:local}, the
expression $e$ must not refer to any previously defined function, but
this is not really a restriction as we can always inline all function
definitions (since the language does not allow for recursive
definitions).

We extend the \MOO{} type system in order to \emph{determine} a set of
expressions that are local in the sense of
Definition~\ref{def:local}. The extension is straightforward and in
the style of Volpano~\cite{volpano-secure-flow}. The record type gets
split into \emph{safe} and \emph{unsafe} records, with type
instantiation that enables to ``cast-down'' safe to unsafe;
complementary, record projection typing gets changed to type as unsafe
record projections explicitly accessing the \PROP{is-installed}
property. The intuition is that functional expressions having
\emph{principal} type with safe record argument are guaranteed not to
access its \PROP{is-installed} property.

Equipped with the above typing machinery, each \MOO{}
sub-expression---no matter where it appears---that have type
$\mathtt{package} \to \mathtt{t}$, for some \texttt{t}, can be tested
for locality as follows:
\begin{enumerate}
\item if it can be typed under the premise that \texttt{u} is a list
  of \emph{safe} packages, then the expression is local
  \begin{enumerate}
  \item if, moreover, its principal type is an arrow from \emph{safe}
    packages to something, the expression is fully determined without
    the candidate solution
  \item otherwise, the expression depends on the property
    \PROP{is-installed} of its sole argument
  \end{enumerate}
\item otherwise the expression is not local
\end{enumerate}

Case (1a) is the luckiest: the sub-expression can be pre-computed on
all packages of the universe, its value stored in a fresh property
name (to be declared in the preamble), and replaced by a field access
the fresh property. Case (1b) requires the additional efforts of
(statically) computing \emph{two} possible values of the
sub-expression, according to the possible values of
\PROP{is-installed}, and of tweaking the program to lookup one or
another fresh property according to the actual \PROP{is-installed}
value at runtime. Case (2) is the worst case, where no partial
evaluation is possible due to non locality.

\begin{example}
  To demonstrate partial evaluation in practice we reconsider
  Example~\ref{ex:freshness}. It contains two expressions having types
  $\mathtt{package}\to\mathtt{t}$ for some \texttt{t}. The first one,
  sub-expression of the \texttt{is-recent} definition body, belong to
  case (1a) (unrelated to solution), while the second one needs the
  property \PROP{is-installed} of its argument, still being
  local. Partial evaluation will rewrite the \MOO{} program leading to
  something like:
  \begin{lstlisting}
let is-recent p = forall (fun q -> q.fresh0) u
maximize cardinality
  (fun p -> if p.is-installed
            then q.fresh1
            else q.fresh2)
  \end{lstlisting}
  where \texttt{fresh0}, \texttt{fresh1}, \texttt{fresh2} are fresh
  properties defined as follows. \texttt{fresh0} will be \texttt{true}
  for all ``most recent packages'', \texttt{fresh1} will inherit from
  \texttt{fresh0}, and \texttt{fresh2} will be the constant
  \texttt{false}.
\end{example}

The limits of the partial evaluation approach are demonstrated by the
example of minimizing multiple installed versions of packages
(Example~\ref{ex:multiple:versions}). In that case the partial
evaluator does not bring any advantage since everything depends on the
final installation status of the packages, and there is no additional
information that can be pre-computed independently of the installation
status of the \emph{other} packages in the universe. A similar case is
the maximization of the number of installed packages that have their
\PROP{recommends} (which is a weak, non-mandatory form of package
dependency) satisfied.

A concluding noteworthy scenario is a reprisal of APT pinning handling
(see Example~\ref{ex:apt-pinning}). No matter how ``strictly'' pinning
gets implemented in \MOO, partial evaluation enables to relax the
requirement that pin priorities reach \MOO{} pre-computed, without
neither implementation  burden, nor performance loss for
dependency solvers. The idea is to store in \MOO{} the rules to assign
pin priorities to packages on the usual basis (origin suite, package
name, package version, \ldots) relying on apposite extra properties
and suitable standard library functions (like regular expression
matching). If pinning assignment is encoded as functions from packages
to integers (and hardly will be otherwise), there is no reason for the
implementing expression to access the \PROP{is-installed} property,
given that pinning rules are static. Hence, the resulting
sub-expressions are local---case (1a)---and will be completely removed
during partial evaluation, returning a CUDF document such as those
assumed by Example~\ref{ex:apt-pinning}.


\section{Conclusion}
\label{sec:outro}

The request to alter the installation of component based software
collections as large as \FOSS{} distributions can have a daunting
number of satisfying answers. To choose the ``best'' solution among
them, state of the art package managers implement ad-hoc heuristics
and offer preference mechanisms of limited expressiveness. In this
paper we presented an architecture to specify user preferences about
\FOSS{} packages which is both independent from specific package
managers or distributions and expressive enough to encode several
preference scenarios. The architecture is composed by a format to
encode upgrade scenarios (CUDF) and by a functional language to encode
user preferences (\MOO).

Future work is planned on several directions. First of all, while
syntax and formal semantics of CUDF have been studied already, various
properties of \MOO{} still need to be investigated in more detail. In
particular we plan to characterize various subsets of \MOO{} that
correspond, after partial evaluation, to language fragments which are
best suited for different encodings of package upgrade problems (SAT,
PBO, constraint programming, etc.).

We also plan to carry partial evaluation further in the direction of
getting rid of data types at partial evaluation stage, so that only
integers (which are the preferred data type for the optimization
community) remain after that.

Finally, the mentioned corpus of upgrade scenarios coming from
different distributions is actually being collected with the final
goal of organizing a recurrent dependency solving
competition. Ideally, such a that forum will become a venue where the
package manager developer community meets the research community on
constraint solving. Both communities could profit from this: package
managers can use complete and more powerful dependency solving tools,
and it gives the research community access to a large corpus of
real-life optimization problems of non-trivial size.

\pagebreak



\begin{thebibliography}{10}

\bibitem{strongdeps-esem-2009}
P.~Abate, J.~Boender, R.~{Di Cosmo}, and S.~Zacchiroli.
\newblock Strong dependencies between software components.
\newblock In {\em Empirical Software Engineering and Measurement 2009}, 2009.
\newblock To appear.

\bibitem{augustsson-patternmatching}
L.~Augustsson.
\newblock Compiling pattern matching.
\newblock In {\em Functional Programming Languages and Computer Architecture},
  volume 201 of {\em LNCS}, pages 368--381. Springer-Verlag, 1985.

\bibitem{eclipse09iwoce}
D.~L. Berre and P.~Rapicault.
\newblock Dependency management for the {E}clipse ecosystem.
\newblock In R.~D. Cosmo and P.~Inverardi, editors, {\em IWOCE 2009 (this
  volume)}, Aug. 2009.

\bibitem{kahn-miniml}
D.~Cl\'{e}ment, T.~Despeyroux, G.~Kahn, and J.~Despeyroux.
\newblock A simple applicative language: mini-ml.
\newblock In {\em Conference on {LISP} and functional programming}, pages
  13--27, New York, 1986. ACM.

\bibitem{milner-principal-type}
L.~Damas and R.~Milner.
\newblock Principal type-schemes for functional programs.
\newblock In {\em ACM Symposium on Principles of Programming Languages (POPL)},
  pages 207--212, USA, 1982. ACM.

\bibitem{mancoosi-wp1d1}
R.~{Di Cosmo} and S.~Cousin.
\newblock Project presentation.
\newblock Deliverable D1.1, The {Mancoosi} project, Jan. 2008.
\newblock \url{http://www.mancoosi.org/deliverables/d1.1.pdf}.

\bibitem{hotswup:package-upgrades}
R.~{Di Cosmo}, P.~Trezentos, and S.~Zacchiroli.
\newblock Package upgrades in {FOSS} distributions: details and challenges.
\newblock In {\em HotSWUp'08}, pages 1--5. ACM, 2008.

\bibitem{edos-package-management}
{EDOS Project}.
\newblock Report on formal management of software dependencies.
\newblock EDOS Project Deliverable D2.1 and D2.2, Mar. 2006.

\bibitem{edos2006ase}
F.~Mancinelli, J.~Boender, R.~D. Cosmo, J.~Vouillon, B.~Durak, X.~Leroy, and
  R.~Treinen.
\newblock Managing the complexity of large free and open source package-based
  software distributions.
\balancecolumns
\newblock In {\em ASE 2006}, pages 199--208, Tokyo, Japan, Sept. 2006. IEEE CS
  Press.

\bibitem{niemeyer-smart}
G.~Niemeyer.
\newblock {Smart} package manager.
\newblock \url{http://labix.org/smart}, 2008.

\bibitem{apt-howto}
G.~{Noronha Silva}.
\newblock {APT} howto.
\newblock \url{http://www.debian.org/doc/manuals/apt-howto/}, 2008.

\bibitem{esr03artofunix}
E.~S. Raymond.
\newblock {\em The Art of UNIX Programming}.
\newblock Addison-Wesley Professional, 1st edition, Oct. 2003.

\bibitem{remy-typeinf-records}
D.~R\'{e}my.
\newblock Type inference for records in natural extension of ml.
\newblock In C.~A. Gunter and J.~C. Mitchell, editors, {\em Theoretical aspects
  of object-oriented programming}, pages 67--95. MIT Press, Cambridge, MA, USA,
  1994.

\bibitem{mancoosi-wp5d1}
R.~Treinen and S.~Zacchiroli.
\newblock Description of the {CUDF} format.
\newblock Deliverable D5.1, The {Mancoosi} project, Nov. 2008.
\newblock \url{http://www.mancoosi.org/deliverables/d5.1.pdf}.

\bibitem{mancoosi-debconf8}
R.~Treinen and S.~Zacchiroli.
\newblock Solving package dependencies.
\newblock In {DebConf8 Proceedings Team}, editor, {\em DebConf 8}, pages
  18--42, Mar del Plata, Argentina, Aug. 2008.
\newblock \url{https://media.debconf.org/dc8/proceedings/proceedings.pdf}.

\bibitem{trezentos07}
P.~Trezentos, R.~{Di Cosmo}, S.~Lauriere, M.~Morgado, J.~Abecasis,
  F.~Mancinelli, and A.~Oliveira.
\newblock {New Generation of Linux Meta-installers}.
\newblock {\em Research Track of FOSDEM 2007}, 2007.

\bibitem{tucker-opium}
C.~Tucker, D.~Shuffelton, R.~Jhala, and S.~Lerner.
\newblock {OPIUM}: Optimal package install/uninstall manager.
\newblock In {\em ICSE '07}, pages 178--188. IEEE Computer Society, 2007.

\bibitem{volpano-secure-flow}
D.~Volpano, C.~Irvine, and G.~Smith.
\newblock A sound type system for secure flow analysis.
\newblock {\em J. Comput. Secur.}, 4(2-3):167--187, 1996.

\end{thebibliography}

\end{document}